\documentclass[fleqn,twoside]{article}
\usepackage{espcrc2}

\usepackage{graphicx}
\usepackage{epsfig}
\usepackage[figuresright]{rotating}

\newcommand{\be}{\begin{equation}}
\newcommand{\ee}{\end{equation}}
\newcommand{\half}{\frac{1}{2}}

\newcommand{\AmS}{{\protect\the\textfont2
  A\kern-.1667em\lower.5ex\hbox{M}\kern-.125emS}}

\title{Lattice QCD with light dynamical quarks\thanks{Combining the
 contributions of F. Farchioni and L. Scorzato.}}

\author{qq+q Collaboration \\[0.5em]
	F. Farchioni\address[Muenster]{Institut f\"ur Theoretische Physik, 
	Universit\"at M\"unster, Wilhelm-Klemm-Str. 9,\\ 
	D-48149 M\"unster, Germany},
	C. Gebert\address[desy]{
        Deutsches Elektronen-Synchrotron DESY,
        Notkestr.\,85, D-22603 Hamburg, Germany},
	I. Montvay\addressmark[desy], L. Scorzato\addressmark[desy].}

\begin{document}

\begin{abstract}
 We report on the simulation of QCD with light dynamical quarks using
 the two-step multi-boson (TSMB) algorithm.
 In an exploratory study with two flavours of quarks at lattice spacing
 about 0.27 $\rm fm$ and with quark mass down to one sixth of the strange
 quark mass eigenvalue spectra and autocorrelations have been studied.
 Here we present results on the volume dependence as well as tests of
 possible algorithmic improvements.
\end{abstract}

\maketitle

\section{INTRODUCTION}
 Predicting the low energy constants of the chiral effective Lagrangian
 from first principles in QCD is an exciting challenge for lattice gauge
 theories.
 This question has attracted a considerable attention in the last
 few years \cite{Heitger:2000ay,Irving:2001vy,Nelson:2001tb}.
 However, existing unquenched simulations are typically done in a region
 where the light quarks are not light enough, in most cases --
 especially with Wilson-type quarks -- with two light quark flavours
 ($u$ and $d$) having masses larger than half of the strange quark mass
 ($m_{ud} > \half m_s$).
 Since the quark mass dependence in the chiral Lagrangian is explicit,
 it is not necessary to perform simulations at the physical value of the
 $u$- and $d$-quark masses, which remains a too hard task for the
 forseeable future.
 Nevertheless, in order to keep the systematic errors under control, 
 three dynamical quarks should be simulated  and, according to \cite{Sharpe:SHORESH},
 the masses of the light ones should satisfy at least
 $m_{ud} \leq \frac{1}{4} m_s$.

 Simulating light dynamical quarks is a difficult task for numerical
 computations because all known algorithms for QCD have a substantial
 slowing down towards small quark masses.
 The present status has been summarized at the Berlin lattice conference
 \cite{Simulations:BERLIN}.
 Inspired by that discussion, in a recent paper \cite{EPJC:paper} we
 parametrized the cost of an unquenched QCD simulation in two different
 ways:
\begin{eqnarray} \label{eq1.01}
&& C = F\; (r_0 m_\pi)^{-z_\pi} \left(\frac{L}{a}\right)^{z_L}
\left(\frac{r_0}{a}\right)^{z_a} \ , 
\\ \label{eq1.02}
&& C_U = F_U\; \left(\frac{m_\pi}{m_\rho}\right)^{-z_{\pi\rho}}
\left(\frac{L}{a}\right)^{z_L} \left(\frac{r_0}{a}\right)^{z_a} \ .
\end{eqnarray}
 Here $r_0$ is the Sommer scale parameter \cite{Sommer:SCALE}, 
 $m_\pi$ the pion mass, $L$ the lattice extension and $a$ the lattice
 spacing.
 The powers $z_{\pi,\pi\rho,L,a}$ and the overall constants $F$, $F_U$
 are empirically determined.
 The value of the constants $F$, $F_U$ depends on the definition of
 ``cost''.
 Of course, when the quark masses are very light and the
 decay $\rho \to \pi\pi$ is possible, the first parametrization is
 preferable.
 Typical values of the parameters can be taken from Ukawa's
 contribution \cite{Simulations:BERLIN}:
 $F_U=5.9 \cdot 10^6$ flop, $z_{\pi\rho}=6$, $z_L=5$, $z_a=2$.
 Other estimates are in reasonable agreement.
 However most of the simulations, whereon such estimates are based,
 were perfomed at quite large quark masses.
 This is a reason of concern because, as experienced already by
 several collaborations \cite{HMC:problems}, at small quark masses
 new phenomena appear, especially for algorithms based on molecular
 dynamics integrators.

 Given the compelling need of simulations at lighter quarks, we
 performed algorithmic studies in the regime of very light quark
 masses  (though on very rough lattices) \cite{EPJC:paper} using the
 two-step multi-boson algorithm \cite{Montvay:TSMB}.
 Here we give an update of the cost analysis.
 We also include a preliminary study of the volume dependence and a
 test of a possible improvement of the algorithm.

\section{COST OF THE SIMULATIONS} \label{sec:costs}
\subsection{Algorithm}
 Let us first briefly summarize the main features of TSMB.
 The quark determinant of $N_f$ degenerate flavours is represented as
\be \label{eq06}
\left|\det(\tilde{Q})\right|^{N_f} \;\simeq\;
\frac{1}{\det P^{(1)}_{n_1}(\tilde{Q}^2)\;
\det P^{(2)}_{n_2}(\tilde{Q}^2)} \ .
\ee
 $\tilde{Q}$ is the hermitean Wilson-Dirac fermion matrix.
 The polynomials $P^{(1)}$ and $P^{(2)}$ satisfy
\be \label{eq07}
\lim_{n_2 \to \infty} P^{(1)}_{n_1}(x)P^{(2)}_{n_2}(x) =
x^{-N_f/2} \ ,\;
x \in [\epsilon,\lambda] \ .
\ee
 The interval $[\epsilon,\lambda]$ covers the spectrum of $\tilde{Q}^2$
 on a typical gauge configuration.
 The first polynomial $P^{(1)}$ of order $n_1$ is a crude approximation
 and is realized by the multi-boson representation.
 The second polynomial $P^{(2)}$ of order $n_2 \gg n_1$ gives a better
 approximation.
 It is taken into account in the updates by a global noisy Metropolis
 correction step.
 Since for fixed $n_2$ (and outside the interval $[\epsilon,\lambda]$)
 the approximation $P^{(1)}P^{(2)}$ is not exact, a final correction
 step is performed by reweighting the gauge configurations which are
 considered for the evaluation of the expectation values.
 This is done with a sufficiently high order polynomial $P^{(2)}_{n_4}$.
 For more details about the algorithm in this context we refer to
 \cite{EPJC:paper}. 

\subsection{Parameters and the definition of cost}
 The Monte Carlo simulations have been done near the $N_t=4$
 thermodynamical cross-over line.
 We tuned the gauge coupling $\beta$ and the hopping parameter $\kappa$
 in order to explore a range of the quark mass going from $2m_s$ to
 $\frac{1}{5} m_s$ while keeping $r_0/a$ roughly constant between 
 $r_0/a=1.7$ and $r_0/a=1.9$, which  implies $a\simeq0.27$ {\rm fm}.
 Most of the analysis is performed on lattices of size $8^3 \cdot 16$
 implying a physical lattice extension $L \simeq 2.2\,{\rm fm}$.
 For the study of finite volume effects we used $12^3 \cdot 24$ and
 $16^4$ lattices.

 The cost of a simulation can be measured in units of the floating point
 operations (flop) which are necessary to produce a new decorrelated
 configuration (starting from a themalized configuration).
 The decorrelation quantified by the integrated autocorrelation length
 $\tau_{int}$ depends on the quantity whose autocorrelation is observed.
 We consider the average plaquette,  the lowest eigenvalue
 of the fermion matrix, the pion correlator at some fixed distance and
 the pion mass.
 Instead of the number of flops the integrated autocorrelations can also
 be quoted by the necessary number of fermion-matrix vector
 multiplications (MVM's).
 We always count mutiplications by the fermion matrix $\tilde{Q}$.
 In this way the dependence on the computer architecture is reduced.
 For TSMB we have the following approximate formula for the total
 amount of MVM's needed for one update cycle:
\begin{equation} \label{form:sweepcycle}
N \;\simeq\; 
6\,(n_1N_\Phi+N_U)+2\,(n_2+n_3)N_{C}+I_G F_G \ .
\end{equation}
 Here $N_\Phi$ is the number of local bosonic sweeps per update cycle, 
 $N_U$ the number of local gauge sweeps, $N_{C}$ the number 
 global Metropolis accept-reject correction steps, and $I_G$ and $F_G$
 give  the number of MVM's and frequency of the global quasi heatbath.
 ($n_3$ is the order of an auxiliary polynomial $P^{(3)}
 $\cite{EPJC:paper}.)
 Of course, the number of MVM's can easily be converted to flops by
 noting that the number of flops per lattice point (in our code) is
 1344. The quantity in (\ref{form:sweepcycle}) is reported for the 
 runs considered here in the columns 7 and 5 of tables 
 \ref{tab:vol} and  \ref{tab:breakup} respectively.

\subsection{Results about the costs}
 In \cite{EPJC:paper} power fits have been given for the dependence
 of the integrated autocorrelations on the quark mass parameter
 $M_r \equiv (r_0 m_\pi)^2$.
 For instance, in case of the average plaquette a good fit is
\begin{eqnarray}
&& \tau_{int}^{plaq} = c_{plaq} M_r^{z_{plaq}} \ ,
\nonumber \\[0.5em] \label{cost:plaquette}
&& c_{plaq}=7.92(68) \ ,\;  z_{plaq}=-2.02(10) \ ,
\end{eqnarray}
 if the cost is counted in MMVM's ($=10^6$ fermion-matrix vector
 multiplications).
 This corresponds to $z_\pi \simeq 4$ in (\ref{eq1.01}).
 A reasonable fit of $\tau_{int}^{plaq}$ can also be found in the form
 (\ref{eq1.02}) with $z_{\pi\rho} \simeq 6$ if the heaviest quark mass
 point is omitted.

\begin{figure}[htb]
\vspace{-1.5em}
\begin{center}
\epsfig{file=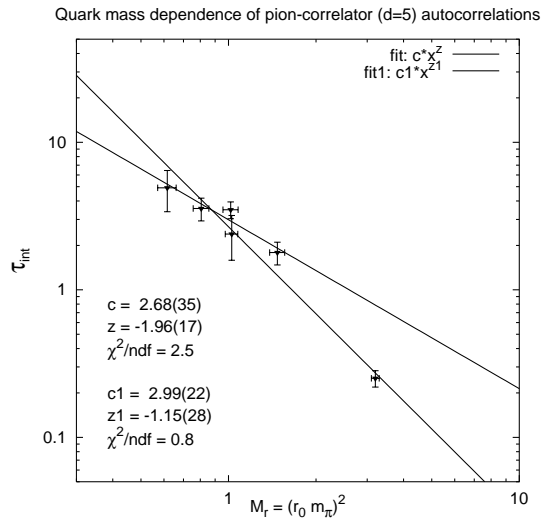,width=7cm,height=10cm,angle=-90}
\end{center}
\vspace{-3.0em}
\caption{\label{taucor5}
 Power fit of the quark mass dependence of the autocorrelation of
 pion correlator at distance 5.}
\vspace{-2.0em}
\end{figure}

\begin{figure}[htb]
\vspace{-0.5em}
\begin{center}
\epsfig{file=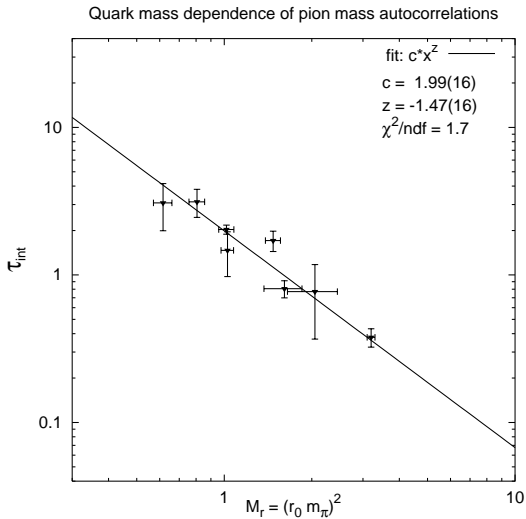,width=7cm,height=10cm,angle=-90}
\end{center}
\vspace{-3.0em}
\caption{\label{taumpi}
 Power fit of the quark mass dependence of the autocorrelation of the
 pion mass.}
\vspace{-0.5em}
\end{figure}

 Another interesting quantity is the pion correlator and/or the pion
 mass.
 For instance, the cost for the pion correlator at timeslice distance
 $d=5$ is shown together with two power fits in figure \ref{taucor5}.
 The fit to all points gives a power similar to $z_{plaq}$:
\begin{equation} \label{cost:picorr}
c_5=2.68(35) \ ,\;  z_5=-1.96(17) \ .
\end{equation}
 However, if the point at $M_r \simeq 3$ is omitted from the fit we
 obtain a lower power:
\begin{equation} \label{cost:picorrr}
c_5=2.99(22) \ ,\;  z_5=-1.15(28) \ .
\end{equation}
 Fitting the cost for obtaining the pion mass, the result is:
\begin{equation} \label{cost:pimass}
c_{m_\pi}=1.99(16) \ ,\;  z_{m_\pi}=-1.47(16) \ .
\end{equation}
 (The data and the fit is shown in figure \ref{taumpi}).
 This corresponds to $z_\pi \simeq 3$ in (\ref{eq1.01}), similarly to
 the result for the minimal eigenvalue of the squared hermitean fermion
 matrix:
\begin{equation} \label{cost:min}
c_{min}=5.36(80) \ ,\;  z_{min}=-1.48(25) \ .
\end{equation}
 It is remarkable that the pion mass has a substantially lower cost
 than, for instance, the average plaquette.
 In fact, at $M_r \simeq 0.5$ the cost ratio is already almost 10 in
 favour of the pion mass.
 This is partly due to the intrinsic fluctuation present in the
 pion propagator which is originated from the freedom of randomly
 choosing the position of the source.

%

\subsection{Analysis with correction factors}
 The importance of the final correction step by the reweighting factors
 depends on the choice of the second polynomial $P^{(2)}$.
 For good enough $P^{(2)}$ the reweighting is negligible.
 In \cite{EPJC:paper} this has been the case for all but one runs.
 (The run where the reweighting was not completely negligible has
 the index (i).)
 Of course, at very small quark masses $P^{(2)}_{n_2}$ becomes good
 enough only for a very large order $n_2$.
 In such a case it is a question whether it would be more advantageous
 to keep $n_2$ smaller and tolerate a non-trivial reweighting.

 The consequences of a poor $P^{(2)}$ at small quark masses could,
 however, be rather unpleasant.
 An example has been shown in \cite{EPJC:paper}: in run $(i)$ at
 $M_r \simeq 0.6$ one of the four parallel runs had a serious
 fluctuation with exceptionally small eigenvalues and, as a consequence,
 very small correction factors in a substantial part of the run (see
 figure 2 in \cite{EPJC:paper}).
 In this part of the run also a few negative eigenvalues of the fermion
 matrix have been observed.
 The smallness of the corresponding reweighting factors implies that
 with a better second polynomial $P^{(2)}$ such fluctuations could have
 been suppressed.

\begin{figure}[htb]
\vspace{-1.0em}
\begin{center}
\epsfig{file=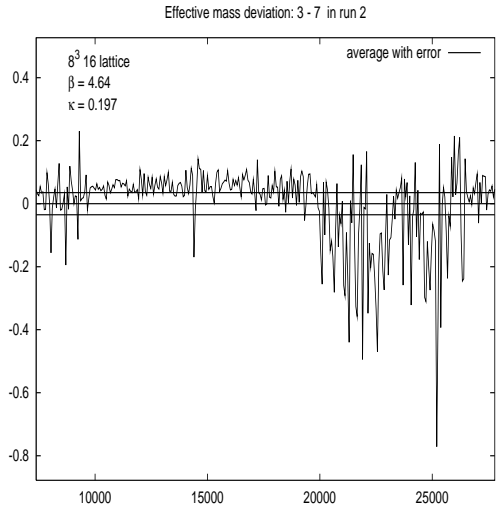,
        width=7cm,height=7cm,angle=-90}
\end{center}
\vspace{-3.0em}
\caption{\label{8c16b464k197.masscorr.3cm7}
 The linearized deviation of the effective pion mass calculated from
 the distance pair 3 and 7 as a function of the number of update
 cycles.}
\vspace{-1.0em}
\end{figure}

 The question is, what is the consequence of such a fluctuation on the
 integrated autocorrelations and on the related magnitude of
 statistical errors.
 This can be investigated, for instance, by considering the linearized
 deviation of effective pion masses which appears in the linearization
 method for determining the autocorrelations of functions of primary
 averages \cite{Frezzotti:BENCHMARK}.
 For an illustration see the case of the effective mass between
 timeslice distances $d=3$ and $d=7$ after including the reweighting
 factors in the above mentionned run of
 \cite{EPJC:paper} (figure \ref{8c16b464k197.masscorr.3cm7}).
 As this figure shows, the exceptionally small eigenvalues and the
 corresponding small reweighting factors are accompanied by a large
 downwards fluctuation of the effective mass $m_\pi^{eff}$.
 This implies that in this run the average value of $m_\pi^{eff}$ is
 smaller than in the ``normal'' runs and at the same time its error is
 larger.
 This pulls down the overall mass estimate: the average of all four
 parallel runs gives for this effective mass
 $m_\pi^{eff} = 0.4160(103)$.
 For the three other runs the result is: $m_\pi^{eff} = 0.4298(63)$.
 Our conclusion is that it is probably better to suppress the kind of
 fluctuations in figure \ref{8c16b464k197.masscorr.3cm7} by
 sufficiently improving $P^{(2)}$.

\section{VOLUME DEPENDENCE} \label{sec:volume}
 The physical volume of the lattices in \cite{EPJC:paper} is such that
 strong finite volume effects are not expected.
 For checking this and for observing the behaviour of our algorithm in
 increasing volumes new simulations were performed on a $12^3 \cdot 24$
 and a $16^4$ lattices (see (f12) and (e16) in table \ref{tab:vol}).
 In order to hold all the quantities fixed, except for the volume,
 run (f12) has the same value of $\beta$ and $\kappa$ as run (f) and run
 (e16) the same as run (e).
 The lattice extensions deduced from $r_0/a$ are collected in the last
 column of table \ref{tab:vol}.
\begin{table*}
\caption{Analysis on larger volumes for run (e) at $\beta=4.76$,
 $\kappa=0.190$ and run (f) at $\beta=4.80$, $\kappa=0.190$.
\label{tab:vol}}
\begin{tabular}{|c|c|c|c|c|c|c|c|c|c|c|}
\hline
run & $n_1$ & $n_2$ & $n_3$  & $\lambda$ & $\epsilon\cdot 10^4$ 
  & $kMVM$ & $\tau_{int}^{plaq}[flop]$ & $\tau_{int}^{m_\pi}[flop]$ & $L[fm]$
   \\ \hline
(e) & 44    & 360   & 380    & 3.6       & 2.7 &
   8.50 & 4.59(37) $\cdot 10^{13}$ & 1.94(31) $\cdot 10^{13}$& 2.31(6)\\ \hline
(e16)&72    & 350   & 440    & 3.6       & 2.7 &
   12.4 & 7.48(1.31) $\cdot 10^{14}$ & 5.02(55) $\cdot 10^{13}$ & 4.57(9)\\\hline
(f) & 44  & 360   & 380    & 3.6       & 2.7 &
   8.48 &7.47(84) $\cdot 10^{13}$ & 1.76(59) $\cdot 10^{13}$ & 2.25(4)\\ \hline
(f12)&72    & 500   & 560   & 3.4       & 1.36 &
   12.3 & 2.40(41) $\cdot 10^{14}$ & 4.52(82) $\cdot 10^{13}$ & 3.02(9)\\\hline
\end{tabular}
\end{table*}

 When going to a larger lattice one has to adjust in TSMB the first
 polynomial order $n_1$.
 Assuming no large finite volume effects the interval given by
 $\epsilon$ and $\lambda$ remains unchanged.
 Therefore it is not needed to change $n_2$ and $n_3$.
 This is true for the runs (e) and (e16).
 For run (f) the lower boundary $\epsilon$ was not optimal
 therefore we used a better (i.e.~smaller) one in run (f12).
 Consequently, $n_2$ and $n_3$ had to be increased there.

 A remarkable feature in table \ref{tab:vol} is that on the larger
 lattices the autocorrelation of the pion mass becomes even more
 favourable compared to the average plaquette.
 This has the consequence that the cost for an integrated
 autocorrelation length of the pion mass increases substantially slower
 than the number of lattice points.
 As expected only a tiny finite-size effect is observed for the physical
 quantities measured.

\section{DETERMINANT BREAKUP} \label{sec:breakup}
 Inspired by \cite{AHasenfratz:breakup} (see also \cite{Hasenbusch:breakup})
 we tried to improve autocorrelations by ``determinant breakup''.
 In our case this means, for instance, that simulating two flavours
 separately should be more effective than two degenerate flavours
 together.
 In general, one can write the determinant as
\be \label{breakup}
\left|\det(\tilde{Q})\right|^{N_f} = 
\left\{\left|\det(\tilde{Q})\right|^{N_f/N_b}\right\}^{N_b} \ ,
\ee
 with an arbitrary positive integer $N_b$ and use the polynomial
 approximation (\ref{eq06}) for the individual factors on the right hand
 side.

 For tests we have chosen run (b) in \cite{EPJC:paper} where
 $M_r \simeq 3$ (close to the strange quark mass).
 Results are shown in table \ref{tab:breakup}.
 (Note that $\tau_{int}^{m_\pi}$ are obtained here from effective
 masses with the linearization method for autocorrelations and errors 
 and therefore are slightly different from \cite{EPJC:paper}.)
 Tests at smaller quark masses and/or in larger volumes will be done
 later.
\begin{table*}
\caption{Run (b) with determinant breakup in two, four and eight pieces.
Here $\beta=5.04$, $\kappa=0.174$, Vol=$8^3\cdot 16$. 
See also \cite{EPJC:paper} for the other quantities measured at this point.
\label{tab:breakup}}
\begin{tabular}{|l|c|c|c|c|c|c|c|}
\hline 
$N_b$ (flavours) & $n_1$ & $n_2$ & $n_3$ & $kMVM$ &
$\tau_{int}^{plaq}[MVM]$ & $\tau_{int}^{min}[MVM]$ & 
$\tau_{int}^{m_\pi}[MVM]$\\ \hline
1 ($N_f=2$) & 28& 90& 120 & 3.22 & 1.13(16) $\cdot 10^6$ &
1.09(19) $\cdot 10^6$ & 3.67(48) $\cdot 10^5$ \\ 
\hline 
2 ($N_f=1+1$)& 20& 84& 100 & 4.45 & 6.05(53) $\cdot 10^5$&
6.45(90) $\cdot 10^5$ & 2.36(36) $\cdot 10^5$ \\ 
\hline 
4 ($N_f=4\times \frac{1}{2}$) & 14 & 72 & 80 & 6.20 & 1.55(37) $\cdot 10^6$
& 1.95(50) $\cdot 10^6$ & 3.60(56) $\cdot 10^5$\\ 
\hline 
8 ($N_f=8\times \frac{1}{4}$) & 10 & 64 & 80 & 9.79 & 1.17(29) $\cdot 10^6$
& 1.63(39) $\cdot 10^6$ & 4.77(58) $\cdot 10^5$\\ 
\hline 
\end{tabular}
\end{table*} 

 Choosing some breakup of the flavours we fixed $n_1$ by requiring a
 constant acceptance rate of $50$-$60\%$.
 The higher polynomial orders $n_2$ and $n_3$ were fixed by keeping
 the relative deviation at the interval ends roughly constant.
 We found that the autocorrelation measured in update cycles 
 is reduced when breaking up the determinant in more pieces.
 However the costs per update cycles are increasing.
 It is remarkable that the cost is rising less than one would expect by
 just looking at the numbers for $n_1$ and $n_2$.
 The reason is that the total number of iterations in the global quasi
 heatbath $I_G$ in (\ref{form:sweepcycle}) are roughly constant,
 although the number of boson fields $N_b n_1$ is increasing.
 Putting everything together we find that at this simulation point one
 can gain from determinant breakup almost a factor of two when splitting
 up two flavours in two pieces.

\section{EIGENVALUES} \label{sec:eigenvalues}
 The study of the low lying modes of the Wilson-Dirac operator ($Q$) is
 interesting both physically and algorithmically.
 It is an interesting question how the qualitative features of the
 spectrum change with increasing volume.
 As described in \cite{EPJC:paper}, we used the implicitely restarted
 Arnoldi method \cite{Maschho:PARPACK} to compute a portion of the
 eigenvalues of the non-hermitean Wilson-Dirac fermion matrix $Q$.
 Typically we computed the eigenvalues with lower real part of the
 matrix $Q$ and those with smallest modulus of the preconditioned matrix
 $\bar{Q}$ (as defined in \cite{EPJC:paper}). 
 By use of the analytical relations in \cite{EPJC:paper} we had access
 to a reasonable part of the spectrum of $Q$ around the origin.

\begin{figure}[htb]
\vspace{-0.5em}
\begin{center}
\epsfig{file=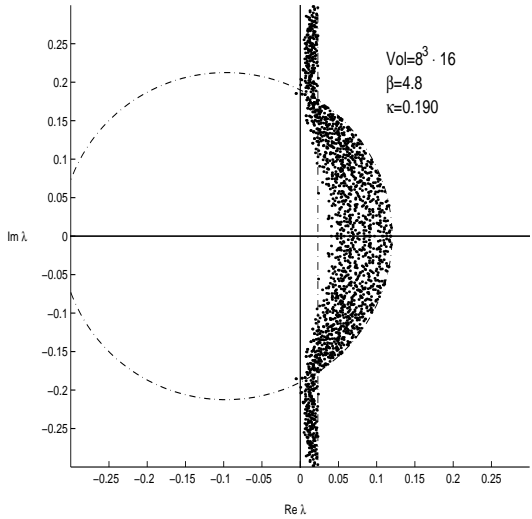,width=7cm,height=7cm,angle=0}
\end{center}
\vspace{-3.0em}
\caption{\label{eigen8_48_190}
 Eigenvalues of the non-hermitean fermion matrix in run (f).}
\vspace{-1.0em}
\end{figure}

\begin{figure}[htb]
\vspace{-0.5em}
\begin{center}
\epsfig{file=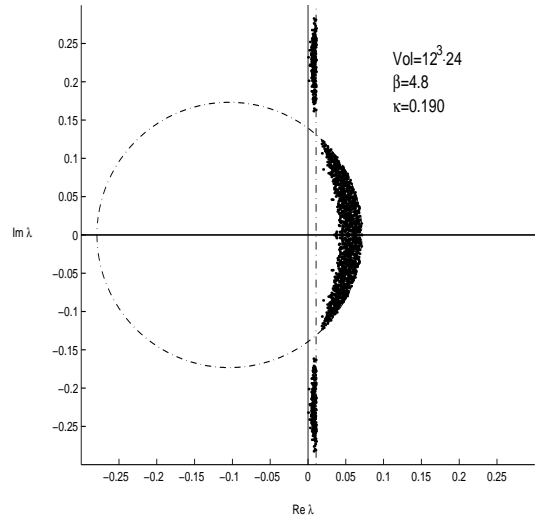,width=7cm,height=7cm,angle=0}
\end{center}
\vspace{-3.0em}
\caption{\label{eigen12_48_190}
 Eigenvalues of the non-hermitean fermion matrix in run (f12).}
\vspace{-0.5em}
\end{figure}

 In figure \ref{eigen8_48_190} we plot $O(200)$ eigenvalues from 10
 configurations thermalized at point (f) on the $8^3 \cdot 16$ lattice.
 The portion of spectrum to which we have access is that inside the
 dashed circle {\em and} to the left of the dashed vertical line.
 In figure \ref{eigen12_48_190} we plot again $O(200)$ eigenvalues from
 10 configurations thermalized at point (f12) on $12^3 \cdot 24$
 lattice, which has the same parameters as (f) -- except for the
 volume.
 As expected, the lowest border of the spectrum does not essentially
 move, even if the density of eigenvalues is higher (and as a
 consequence, the portion of spectrum that we can cover is smaller).

\section{CONCLUSION}
 In conclusion, the cost increase with the lattice volume is quite
 acceptable because it is close to the trivial volume factor.
 In case of the autocorrelation of the pion mass the observed increase
 turns out to be even smaller.

 The determinant breakup is an easy and effective method to speed up
 the gauge field update.



\begin{thebibliography}{10}

\bibitem{Heitger:2000ay}
J.~Heitger, R.~Sommer and H.~Wittig, ALPHA Collaboration,
\newblock Nucl.\ Phys.\ B {\bf 588} (2000) 377, hep-lat/0006026.

\bibitem{Irving:2001vy}
A.~C.~Irving {\em et~al.}, UKQCD Collaboration,
\newblock Phys.\ Lett.\ B {\bf 518} (2001) 243, hep-lat/0107023.

\bibitem{Nelson:2001tb}
D.~R.~Nelson, G.~T.~Fleming and G.~W.~Kilcup,
\newblock hep-lat/0112029.

\bibitem{Sharpe:SHORESH}
S.~R. Sharpe and N.~Shoresh,
\newblock Phys. Rev. {\bf D62} (2000) 094503, hep-lat/0006017.

\bibitem{Simulations:BERLIN}
Contributions of N.~H. Christ, S.~Gottlieb, K.~Jansen, Th.~Lippert,
A.~Ukawa, H.~Wittig,
\newblock Nucl. Phys. Proc. Suppl. {\bf 106} (2002).

\bibitem{EPJC:paper}
F.~Farchioni, C.~Gebert, I.~Montvay and L.~Scorzato,
\newblock hep-lat/0206008, to appear on Eur. Phys. J. C.

\bibitem{Sommer:SCALE}
R.~Sommer,
\newblock Nucl. Phys. {\bf B411} (1994) 839, hep-lat/9310022.

\bibitem{HMC:problems}
UKQCD Collaboration, B. Joo et al.,
\newblock Nucl. Phys. Proc. Suppl. {\bf 106} (2002) 1073, 
hep-lat/0110047.

\bibitem{Montvay:TSMB}
I.~Montvay,
\newblock Nucl. Phys. {\bf B466} (1996) 259, hep-lat/9510042.

\bibitem{Frezzotti:BENCHMARK}
ALPHA, R.~Frezzotti, M.~Hasenbusch, U.~Wolff, J.~Heitger, and K.~Jansen,
\newblock Comput. Phys. Commun. {\bf 136} (2001) 1, hep-lat/0009027.

\bibitem{AHasenfratz:breakup}
A.~Hasenfratz and A.~Alexandru,
\newblock Phys.\ Rev.\ D {\bf 65} (2002) 114506,
hep-lat/0203026; hep-lat/0207014.

\bibitem{Hasenbusch:breakup}
M.~Hasenbusch,
\newblock Phys.\ Rev.\ D {\bf 59} (1999) 054505,
hep-lat/9807031.

\bibitem{Maschho:PARPACK}
K.~Maschhoff and D.~C. Sorensen (1996),\\
\newblock {\small http://www.caam.rice.edu/software/ARPACK/.}

\end{thebibliography}
\end{document}